\documentclass[11pt,a4paper]{article}

\usepackage{indentfirst}
\usepackage{graphicx}
\usepackage{epsfig}

\usepackage{latexsym}
\usepackage{amsmath}
\usepackage{amssymb}
\usepackage{amsfonts}
\usepackage{mathrsfs}

\usepackage{morefloats}

\usepackage{natbib}

\newtheorem{lemma}{Lemma}
\newtheorem{corollary}{Corollary}

\providecommand\bnabla{\boldsymbol{\nabla}}
\providecommand\bcdot{\boldsymbol{\cdot}}

\begin{document}

\begin{center}
{
    \Large\textbf{A THEORETICAL FRAMEWORK OF VORTICITY DYNAMICS FOR TWO DIMENSIONAL FLOWS ON FIXED SMOOTH SURFACES}
    \footnote{Corresponding Author: XIE XiLin, Department of Mechanics \& Engineering Science, Fudan University.
              HanDan Road 220, Shanghai 200433, China. Tel: 0086-21-55664283;
              Email: xiexilin@fudan.edu.cn.}
}
\end{center}

\begin{center}
{   \small
    XIE XiLin\\ 
    Department of Mechanics \& Engineering Science, Fudan University, Shanghai 200433, China.\\
}
\end{center}

\begin{center} \textbf{Original Manuscript updated on {\today} }\end{center}

\begin{abstract}
Two dimensional flows on fixed smooth surfaces have been studied in the point of view of vorticity dynamics. Firstly,
the related deformation theory including kinematics and kinetics is developed. Secondly, some primary relations in vorticity dynamics have been extended to two dimensional flows on fixed smooth surface through which a theoretical
framework of vorticity dynamics have been set up, mainly including governing equation of vorticity, Lagrange theorem
on vorticity, Caswell formula on strain tensor and stream function \& vorticity algorithm with pressure Possion equation for incompressible flows. The newly developed theory is characterized by the appearances of surface curvatures in some primary relations and governing equations.
\end{abstract}


\section{Introduction}
Two dimensional flows/deformations are naturally referred to flows/deformations of continuous mediums whose
geometrical configurations can be taken as two dimensional surfaces embedded in three dimensional Euclidian space.
Physically, the values of the thickness of continuous mediums are quite smaller than the ones of character scales
along flow directions. Such flows/deformations may be divided into two groups. The first one is termed as
\emph{flows on fixed surfaces} that can be considered as models of long wandering rivers on the earth, real floods
spreading over plains, depressions and valleys, atmosphere motions on planets, creeping flows on curved surfaces, flat
soap films and so on. The second one is termed as \emph{self-motions}/\emph{deformations} as models of bubble, cell
and capsule deformations, vibrations of solid membranes, oil contaminations on sea surfaces et al.

The ways of attaining governing equations of natural conservation laws can also be separated into two
kinds. The first kind is to do approximations based on general full dimensional governing equations. For example,
\cite{Roberts-2006} attained models of thin fluid flows with inertia on curved substrates based on central manifold
approximation. The second kind is to do analysis in the point of view of general continuum mechanics. The equations
governing fluid motion in a surface or interface have been studied by \cite{Aris-1962}. Very recently, a novel kind of
finite deformation theory of continuous mediums whose geometrical configurations are two dimensional smooth surfaces
have been developed by \cite{XXL-2013}. It mainly includes the definitions of initial and current configurations,
deformation gradient tensor with its primary properties, deformation descriptions, transport theories and governing
equations corresponding to conservation laws. The general theory is suitable for any kind of two dimensional
flows/deformations. Particularly, the present study puts focus on flows on fixed smooth surface in the point of view
of vorticity dynamics \cite[see][]{WJZ-2005}.

\section{Finite Deformation Theory}

\subsection{Configurations, Coordinates, Curvatures \& Field Operators}

\begin{figure}
\begin{center}
    \includegraphics[height=60mm]{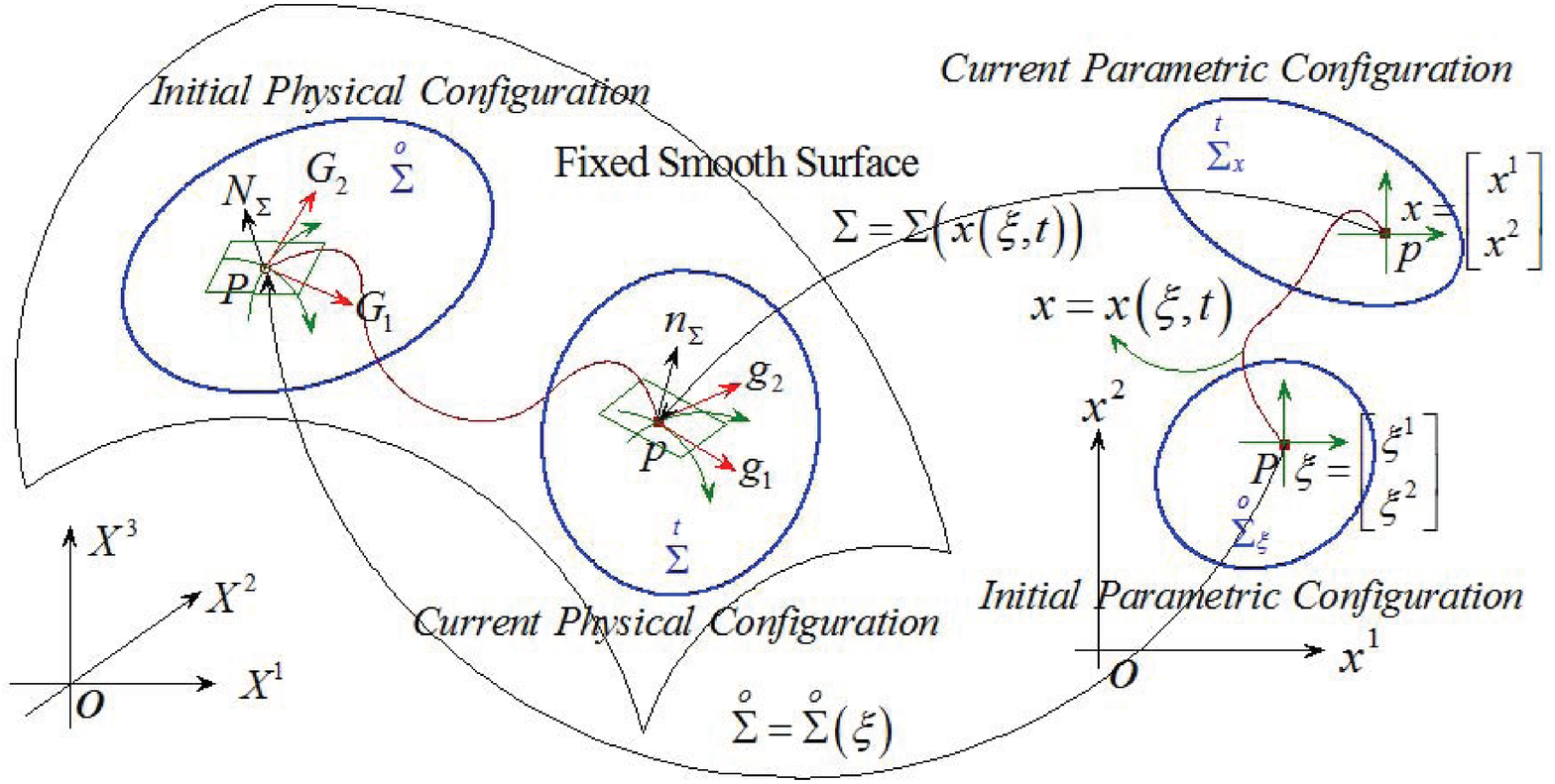}
\end{center}
\caption{Sketch of the initial/current and physical/parametric configurations, where $\mathbf{n}_\Sigma$}
         and $\mathbf{N}_\Sigma$ denote the surface normal vectors associated to the current and initial physical configurations respectively, $\{\mathbf{g}_i\}^2_{i=1}$ and $\{\mathbf{G}_A\}^2_{A=1}$ are local co-variant bases. The tangent space denoted by $\boldsymbol{T\Sigma}$ is span by the co- or contra-variant basis.
\label{Draw1}
\end{figure}

A two dimensional flow on a fixed smooth surface is sketched in Figure \ref{Draw1}. The surface is represented by the
vector valued mapping $\boldsymbol{\Sigma}=\boldsymbol{\Sigma}(x)\in\mathbb{R}^3$, and $\{x^i\}^2_{i=1}$ is the parametric coordinates. The flow/deformation can be represented by the smooth diffeomorphism
$x=x(\xi,t)\in\mathscr{C}^\infty(\overset{\circ}{\Sigma}_\xi,\overset{t}{\Sigma}_x)$ in the parametric space, where
$\{x^i\}^2_{i=1}$ and $\{\xi^A\}^2_{A=1}$ denote Eulerian and Lagrangian coordinates respectively, and $\overset{\circ}{\Sigma}_\xi$ and $\overset{t}{\Sigma}_x$ are termed as the initial and current parametric configurations respectively.
Correspondingly, $\overset{\circ}{\Sigma}:=\boldsymbol{\Sigma}(\overset{\circ}{\Sigma}_\xi)$ and
$\overset{t}{\Sigma}:=\boldsymbol{\Sigma}(\overset{t}{\Sigma}_x)$ are termed as the initial and current physical
configurations respectively. The smooth surface can be taken as a two dimensional Riemannian manifold with the fundamental quantity
of the first kind $\{g_{ij}\triangleq(\boldsymbol{g}_i,\boldsymbol{g}_j)\}^2_{i,j=1}$, and the one of the second kind
$\{b_{ij}\triangleq(\frac{\partial\boldsymbol{g}_j}{\partial x^i}(x),\boldsymbol{n}_\Sigma)\}^2_{i,j=1}$. Gaussian curvature and mean curvature are defined as $K_G\triangleq \det[b_{ij}]/\det[g_{ij}]$ and $H\triangleq b^s_s$ respectively. Einstein summation convention is adopted throughout the paper with subscripts and superscripts representing co- and contra-variant quantities respectively, and quantities with capital/small scripts associate with initial/current configurations.

Generally, the \emph{surface gradient operator} $\overset{\Sigma}{\bnabla}\equiv \boldsymbol{g}^l\frac{\partial}{\partial x^l}$ is defined as, say $\mathbf{\Phi}\in\mathscr{T}^2(\boldsymbol{T\Sigma})$,
\begin{eqnarray}
    \overset{\Sigma}{\bnabla}\circ-\mathbf{\Phi}
  &&\equiv\left(\boldsymbol{g}^l\frac{\partial}{\partial x^l}\right)\circ-
    \left( \Phi^i_{\cdot j}\boldsymbol{g}_i\otimes\boldsymbol{g}^j \right)
    \triangleq \boldsymbol{g}^l\circ-
    \frac{\partial}{\partial x^l}\left( \Phi^i_{\cdot j}\boldsymbol{g}_i\otimes\boldsymbol{g}^j \right)
    \nonumber \\
  &&=\nabla_l\Phi^i_{\cdot j}(\boldsymbol{g}^l\circ-\boldsymbol{g}_i)\otimes\boldsymbol{g}^j
    +\Phi^i_{\cdot j}b_{li}(\boldsymbol{g}^l\circ-\boldsymbol{n})\otimes\boldsymbol{g}^j
    +\Phi^i_{\cdot j}b_l^j(\boldsymbol{g}^l\circ-\boldsymbol{g}_i)\otimes\boldsymbol{n}
    \nonumber
\end{eqnarray}
where $\circ-$ can be any available algebra tensor operator, $\nabla_l$ denotes the co-variant derivative/differentation of the tensor component, e.g.
\begin{equation*}
    \nabla_l\Phi^i_{\cdot j}\triangleq
    \frac{\partial\Phi^i_{\cdot j}}{\partial x^l}(x,t)+\Gamma^i_{ls}\Phi^s_{\cdot j}
    -\Gamma^s_{lj}\Phi^i_{\cdot s}
\end{equation*}
where $\Gamma^i_{ls}$ denotes Christoffel symbol of the second kind. The contra-variant derivative relates generally to the co-variant one through $\nabla^l\triangleq g^{lt}\nabla_t$.
On the other hand, one defines the so termed \emph{Levi-Civita connection operator} $\bnabla\equiv \boldsymbol{g}^l\nabla_\frac{\partial}{\partial x^l}$
\begin{equation*}
    \bnabla\circ-\mathbf{\Phi}\equiv(\boldsymbol{g}^l\nabla_{\frac{\partial}{\partial x^l}})\circ-
    (\Phi^i_{\cdot j}\boldsymbol{g}_i\otimes\boldsymbol{g}^j)\triangleq\boldsymbol{g}^l\circ-\nabla_{\frac{\partial}{\partial x^l}}(\Phi^i_{\cdot j}\boldsymbol{g}_i\otimes\boldsymbol{g}^j)
    =\nabla_l\Phi^i_{\cdot j}(\boldsymbol{g}^l\circ-\boldsymbol{g}_i)\otimes\boldsymbol{g}^j
\end{equation*}
where $\nabla_{\frac{\partial}{\partial x^l}}$ denotes Levi-Civita connection defined on the surface/Riemannian manifold \cite[see][]{Durovin-1992}. It should be noted that the change of the order of co- and contra-variant derivatives must be related to Riemannian-Christoffel tensor that can be represented by Gaussian curvature and metric tensor for two dimensional Riemannian manifolds as revealed by the relation
\begin{eqnarray}
    \nabla_p\nabla^q\Phi^i_{\cdot j}
 &&=\nabla^q\nabla_p\Phi^i_{\cdot j}+R^{i\,\cdot\cdot\,q}_{\cdot sp\,\cdot}\Phi^s_{\cdot j}+R^{\cdot s\cdot q}_{j\cdot p\,\cdot}\Phi^i_{\cdot s}
   \nonumber \\
 &&=\nabla^q\nabla_p\Phi^i_{\cdot j}
   +K_G(\delta^i_p \delta^q_s-g_{sp}g^{iq})\Phi^s_{\cdot j}
   +K_G(g_{jp}g^{sq}-\delta^s_p \delta_j^q)\Phi^i_{\cdot s}
   \nonumber
\end{eqnarray}
where $R^{i\,\cdot\cdot\,q}_{\cdot sp\,\cdot}\triangleq b^i_p b^q_s-b_{sp}b^{iq}$ denotes the component of Riemannian-Christoffel tensor.

It is worthy of note that Levi-Civita connection operator is just valid/effective for surface tensor fields that just have components on the tangent plane, generally denoted by $\mathscr{T}^r(\boldsymbol{T\Sigma})$ where $r\in\mathbb{N}$ denotes the order. However, the surface gradient operator can be applied to arbitrary tensor fields defined on the surface due to its definition is based on differential calculus of tensor normed space.

\subsection{Kinematics}
Generally, the velocity of a fluid partial is determined as follows
\begin{equation}
    \boldsymbol{V}\triangleq\frac{\partial\boldsymbol{\Sigma}}{\partial t}(\xi,t)
   =\frac{\partial x^i}{\partial t}(\xi,t)\frac{\partial\boldsymbol{\Sigma}}{\partial x^i}(x)=:V^i\boldsymbol{g}_i
   \label{Vel}
\end{equation}
Subsequently, the acceleration is
\begin{equation}
    \boldsymbol{a}\triangleq\frac{\partial\boldsymbol{V}}{\partial t}(\xi,t)
   =\left( \frac{\partial V^l}{\partial t}(x,t)+V^s\nabla_sV^l \right)\boldsymbol{g}_l
   +\left( b^{ij}V_i V_j \right )\mathbf{n}
   =:a^l\boldsymbol{g}_l+a_n\boldsymbol{n}
   \label{Acc}
\end{equation}
The material derivative of any surface tensor field is represented by
\begin{equation}
    \frac{\mathrm{d}\mathbf{\Phi}}{\mathrm{d}t}\equiv\dot{\mathbf{\Phi}}\triangleq\frac{\partial\mathbf{\Phi}}{\partial t}(\xi,t)
   =\frac{\partial\mathbf{\Phi}}{\partial t}(x,t)+V^s\frac{\partial\mathbf{\Phi}}{\partial x^s}(x,t),
   \quad\forall\,\mathbf{\Phi}\in\mathscr{T}^r(\boldsymbol{T\Sigma})
   \label{Material Derivative}
\end{equation}
based on the velocity representation \eqref{Vel}.
The deformation gradient tensor denoted by $\mathbf{F}$ is introduced by the relation
\begin{equation}
    \boldsymbol{\Sigma}(x(\xi+\delta\xi,t),t)-\boldsymbol{\Sigma}(x(\xi,t),t)\doteq
    \boldsymbol{\mathbf{F}}\bcdot\left( \overset{\circ}{\boldsymbol{\Sigma}}(\xi+\delta\xi)-\overset{\circ}{\boldsymbol{\Sigma}}(\xi) \right),\quad
    \mathbf{F}:=\frac{\partial x^i}{\partial \xi^A}(\xi,t)\boldsymbol{g}_i(x)\otimes\boldsymbol{G}^A(\xi)
    \nonumber
\end{equation}
All kinds of deformation descriptions can be represented by the deformation gradient tensor with its properties as indicated in \cite{XXL-2013}.

Similar to the familiar Helmholtz velocity decomposition, one has the relation
\begin{eqnarray}
  &&\boldsymbol{V}(x+\delta x,t) - \boldsymbol{V}(x,t)
  \doteq( \boldsymbol{V}\otimes\bnabla )\bcdot\delta\overset{t}{\boldsymbol{\Sigma}} \nonumber \\
  &&=
    \left[\frac{1}{2}(\nabla_jV_i+\nabla_iV_j)\boldsymbol{g}^i\otimes\boldsymbol{g}^j\right]\bcdot\delta\overset{t}{\boldsymbol{\Sigma}}
   +\left[\frac{1}{2}(\nabla_jV_i-\nabla_iV_j)\boldsymbol{g}^i\otimes\boldsymbol{g}^j\right]\bcdot\delta\overset{t}{\boldsymbol{\Sigma}}
   \nonumber\\
   &&+(V^sb_{st}\boldsymbol{g}^t)\bcdot\delta\overset{t}{\boldsymbol{\Sigma}},
   \quad\delta\overset{t}{\boldsymbol{\Sigma}}=\delta x^s\boldsymbol{g}_s(x)\in\boldsymbol{T\Sigma}
   \label{VelDiff}
\end{eqnarray}
Accompanying \eqref{VelDiff} with the dual relations
\begin{equation}
    (\nabla_jV_i-\nabla_iV_j)=:\epsilon_{ji3}\omega^3
    \quad\Leftrightarrow\quad
    \omega^3:=\epsilon^{3ij}\nabla_iV_j
    \label{DualRel}
\end{equation}
the vorticity is defined as $\boldsymbol{\omega}\triangleq\epsilon^{3kl}\nabla_k V_l\,\boldsymbol{n}_\Sigma=\omega^3\boldsymbol{n}_\Sigma$, where $\epsilon_{ji3}:=[\boldsymbol{g}_j,\boldsymbol{g}_i,\boldsymbol{n}_\Sigma]$ and $\epsilon^{3kl}:=[\boldsymbol{n}_\Sigma,\boldsymbol{g}^k,\boldsymbol{g}^l]$ are termed as co- and contra-variant components of Eddington tensor that as the same to metric tensor are insensitive to co- and contra-variant derivatives. And one has the novel velocity decomposition.
\begin{lemma}[Velocity Decomposition]
\begin{equation*}
    \boldsymbol{V}(x+\delta x,t)-\boldsymbol{V}(x,t)
   =\mathbf{D}\bcdot\delta\overset{t}{\boldsymbol{\Sigma}}+\frac{1}{2}\boldsymbol{\omega}\times\delta\overset{t}{\boldsymbol{\Sigma}}
   +\boldsymbol{V}\bcdot\mathbf{K}\bcdot\delta\overset{t}{\boldsymbol{\Sigma}},
\end{equation*}
where $\mathbf{D}:=\frac{1}{2}(\nabla_iV_j+\nabla_jV_i)\boldsymbol{g}^i\otimes\boldsymbol{g}^j$ can be defined as strain tensor, $\mathbf{K}:=b_{ij}\boldsymbol{g}^i\otimes\boldsymbol{g}^j$ is termed as surface curvature tensor. The last appended term is due to the contribution of the surface curvature.
\end{lemma}

\subsection{Kinetics}

Basically, so termed \emph{intrinsic generalized Stokes formula of the second kind} is introduced
\begin{equation}
    \oint_{\partial\Sigma} \boldsymbol{n}\circ-\mathbf{\Phi}dl
   =\int_\Sigma \left(\, \overset{\Sigma}{\bnabla}\circ-\mathbf{\Phi}+H\boldsymbol{n}_\Sigma\circ-\mathbf{\Phi} \,\right)\:\mathrm{d}\sigma
   \label{2nd Stokes formula}
\end{equation}
where $\boldsymbol{n}:=\boldsymbol{\tau}\times\boldsymbol{n}_\Sigma$ is the normal vector of the boundary on the tangent plane in which $\boldsymbol{\tau}$ is the unit tangent vector of the boundary, $\mathbf{\Phi}$ can be any kind of tensor field defined on the surface. It is worthy of note that most of the governing equations for flows on surfaces and the ones for thin enough shells and plates, as indicated by \cite{Aris-1962} and \cite{Chien-1941} respectively, can be directly and readily deduced through \eqref{2nd Stokes formula}. Its detailed proof can be referred to the paper by \cite{XXL-2013}.

On the mass conservation, it can be represented and analysed
\begin{equation}
0=
 \int_{\overset{t}{\Sigma}} \frac{\partial\rho}{\partial t}(x,t)\:\mathrm{d}\sigma
+\oint_{\partial\overset{t}{\Sigma}} \boldsymbol{n}\bcdot(\rho\boldsymbol{V})\:\mathrm{d}l
=\int_{\overset{t}{\Sigma}} \frac{\partial\rho}{\partial t}(x,t)\:\mathrm{d}\sigma
+\int_{\overset{t}{\Sigma}} \bnabla\bcdot(\rho\boldsymbol{V})\:\mathrm{d}\sigma
\nonumber
\end{equation}
where $\rho$ denotes the surface density. Subsequently, the differential form for mass conservation is deduced
\begin{equation}
    \frac{\partial\rho}{\partial t}(x,t)+\bnabla\bcdot(\rho\boldsymbol{V})
   =\left[ \frac{\partial\rho}{\partial t}(x,t)+V^i\frac{\partial\rho}{\partial x^i}(x,t) \right] +
    \rho\nabla_iV^i
   =\dot{\rho}+\rho\,\theta=0,
   \quad \theta:=\nabla_iV^i
   \label{MassCon_Diff}
\end{equation}

On the momentum conservation, it is considered that the rate of change of the momentum are contributed by the surface
tension, inner pressure, inner friction and surface force, in other words it is keeping valid
\begin{equation}
 \int_{\overset{t}{\Sigma}} \frac{\partial(\rho\boldsymbol{V})}{\partial t}(x,t)\:\mathrm{d}\sigma
+\oint_{\partial\overset{t}{\Sigma}} \boldsymbol{n}\bcdot(\rho\boldsymbol{V})\boldsymbol{V}\:\mathrm{d}l
=\int_{\overset{t}{\Sigma}}\rho\, \boldsymbol{a}\:\mathrm{d}\sigma
=\boldsymbol{F}_{ten}+\boldsymbol{F}^{int}_{pre}+\boldsymbol{F}_{vis}+\boldsymbol{F}_{sur}
 \nonumber
\end{equation}
where $\boldsymbol{F}_{ten}:=\oint_{\partial\overset{t}{\Sigma}} \gamma\boldsymbol{n}\:\mathrm{d}l$,
$\boldsymbol{F}^{int}_{pre}:=-\oint_{\partial\overset{t}{\Sigma}} p\boldsymbol{n}\:\mathrm{d}l$,
$\boldsymbol{F}_{vis} :=\oint_{\partial\overset{t}{\Sigma}} \mu\boldsymbol{n}\bcdot
\left[ \left( \nabla_j V_i+\nabla_i V_j\right)
\boldsymbol{g}^i\otimes\boldsymbol{g}^j\right ]\:\mathrm{d}l$
are represented originally as curve integrals that can be transformed directly to surface integrals through the
intrinsic generalized Stokes formula of the second kind, $\gamma$ and $\mu$ denote the coefficients of surface tension
and inner friction/viscousity respectively. $\boldsymbol{F}_{sur}:=\int_{\overset{t}{\Sigma}} \boldsymbol{f}_{sur}
\mathrm{d}\sigma$ is represented directly as surface integral, $\boldsymbol{f}_{sur}$ can be the surface densities of friction, gravity, electromagnetic force and stochastic force et al. As reviewed by \cite{Boffetta-2012} and
\cite{Bouchet-2012}, the surface force usually plays the important role in dynamics of two dimensional flows. Furthermore, the component equations of momentum conservation can be deduced
\begin{equation}
\left\{
\begin{array}{l}
\displaystyle
\rho a_l
 =-\frac{\partial p}{\partial x^l}(x,t)
 +\mu\left[
 g^{ij}\nabla_i \nabla_j V_l
 +{\nabla_l}\left(\nabla^s V_s\right)+K_G V_l
 \right]+f_{sur,l}\\[12pt]
\displaystyle
\rho a_n
=H(\gamma-p)+\mu\left[ 2b^{ij} \nabla_i V_j \right]+f_{sur,n}
\end{array}
\right.
\label{NSE-com}
\end{equation}
where $a_l$ and $a_n$ are the components of acceleration on the tangent plane and in the normal direction
respectively, as shown in \eqref{Acc}. It is revealed by \eqref{NSE-com} that Gaussian curvature accompanying with the tangent velocity takes part in the momentum conservation on the tangent space, however mean curvature accompanying with the surface tension and inner pressure does the contribution to the momentum conservation in the surface normal direction.

On the moment of momentum conservation, one has the general conclusion
\begin{lemma}[Moment of momentum Conservation]
Generally, the mechanical action imposed on the boundary can be represented by the so termed surface stress tensor
\begin{equation*}
    \mathbf{t}=t^i_{\cdot j}\boldsymbol{g}_i\otimes\boldsymbol{g}^j+t^i_{\cdot 3}\boldsymbol{g}_i\otimes\boldsymbol{n}_\Sigma
\end{equation*}
then the moment of momentum conservation takes the form
\begin{equation*}
    \boldsymbol{g}^l\bcdot\left( \mathbf{t}\times\boldsymbol{g}_l \right)+\boldsymbol{m}_\Sigma=0\in\mathbb{R}^3
\end{equation*}
where $\boldsymbol{m}_\Sigma$ denotes the surface force couple.
\end{lemma}

\noindent\textbf{Proof}:
Based on the surface stress tensor, the momentum conservation can be represented generally as
\begin{equation}
    \int_{\overset{t}{\Sigma}} \rho\,\boldsymbol{a}\:\mathrm{d}\sigma
   =\oint_{\partial\overset{t}{\Sigma}} \boldsymbol{n}\bcdot\mathbf{t}\:\mathrm{d}l
   +\int_{\overset{t}{\Sigma}} \boldsymbol{f}_\Sigma\:\mathrm{d}\sigma
   \quad\Rightarrow\quad
   \rho\,\boldsymbol{a}=\overset{\Sigma}{\nabla}\bcdot\mathbf{t}+\boldsymbol{f}_\Sigma
   \label{Momentum Equ.}
\end{equation}
On the other hand, the moment of momentum conservation is represented as
\begin{equation*}
    \int_{\overset{t}{\Sigma}} \rho\,\boldsymbol{a}\times\boldsymbol{\Sigma}\:\mathrm{d}\sigma
   =\oint_{\partial\overset{t}{\Sigma}} (\boldsymbol{n}\bcdot\mathbf{t})\times\boldsymbol{\Sigma}\:\mathrm{d}l
   +\int_{\overset{t}{\Sigma}} \boldsymbol{f}_\Sigma\times\boldsymbol{\Sigma}\:\mathrm{d}\sigma
   +\int_{\overset{t}{\Sigma}} \boldsymbol{m}_\Sigma\:\mathrm{d}\sigma
\end{equation*}
with the differential form
\begin{equation}
    \rho\,\boldsymbol{a}\times\boldsymbol{\Sigma}=\overset{\Sigma}{\bnabla}\bcdot(\mathbf{t}\times\boldsymbol{\Sigma})
    +\boldsymbol{f}_\Sigma\times\boldsymbol{\Sigma}+\boldsymbol{m}_\Sigma
    =\left[\, \left( \overset{\Sigma}{\bnabla}\bcdot\mathbf{t} \right)\times\boldsymbol{\Sigma}+\boldsymbol{g}^l\bcdot(\mathbf{t}\times\boldsymbol{g}_l)\,\right]
    +\boldsymbol{f}_\Sigma\times\boldsymbol{\Sigma}+\boldsymbol{m}_\Sigma
   \label{Moment of Momentum Equ.}
\end{equation}
Substituting \eqref{Momentum Equ.} into \eqref{Moment of Momentum Equ.}, the proof is completed. Furthermore, thanks to the representation
\begin{equation*}
    \boldsymbol{g}^l\bcdot(\mathbf{t}\times\boldsymbol{g}_l)
   =-t^{ij}\epsilon_{ij3}\boldsymbol{n}_\Sigma
   +\sqrt{g}(-t^2_{\cdot 3}\boldsymbol{g}^1+t^1_{\cdot 3}\boldsymbol{g}^2),
   \quad g:=\det[g_{ij}]
\end{equation*}
it can be concluded that
\begin{corollary}[On Representations of Surface Stress Tenor]
    The symmetry of the components of surface stress tensor on the tangent space, i.e. $t_{ij}=t_{ji}$, corresponds to the vanishing of the component of surface force couple in the surface normal direction. And the appearance of surface stress tensor in the surface normal direction, i.e. $t^i_{\cdot 3}\neq 0$, corresponds to the existence of components of surface force couple on the tangent space.
\end{corollary}

As an application, the stress tensor corresponding to the actions of surface tension, inner pressure,
inner friction as discussed previously takes the form
\begin{equation}
    \mathbf{t}=(\gamma-p)\mathbf{I}
    +\mu\left( \nabla_j V_i+\nabla_i V_j \right)\boldsymbol{g}^i\otimes\boldsymbol{g}^j,
    \quad \mathbf{I}:=g_{ij}\boldsymbol{g}^i\otimes\boldsymbol{g}^j
    \label{Constitutive Relation}
\end{equation}
that corresponds to the case of full zero surface force couple as usually satisfied by Newtonian fluid flows.

On energy conservation, one has the identity
\begin{eqnarray}
    \int_{\overset{t}{\Sigma}}\,\frac{\partial}{\partial t}\left( e+\frac{|\boldsymbol{V}|^2}{2} \right)(x,t)\:\mathrm{d}\sigma
   +\oint_{\partial\overset{t}{\Sigma}}\,\boldsymbol{n}\bcdot(\rho\boldsymbol{V})\left( e+\frac{|\boldsymbol{V}|^2}{2} \right)\:\mathrm{d}l
   \nonumber \\
   =\oint_{\partial\overset{t}{\Sigma}}\,(\boldsymbol{n}\bcdot\mathbf{t})\bcdot\boldsymbol{V}\:\mathrm{d}l
   +\int_{\overset{t}{\Sigma}}\,\boldsymbol{f}_\Sigma\bcdot\boldsymbol{V}\:\mathrm{d}\sigma
   +\int_{\overset{t}{\Sigma}}\,q_\Sigma\:\mathrm{d}\sigma
   \nonumber
\end{eqnarray}
where $e$ denotes the surface density of internal energy and $q_\Sigma$ the heat flux. Furthermore, using \eqref{Momentum Equ.}, one can arrive at
\begin{eqnarray}
    \rho\,\dot{e}=\boldsymbol{g}^l\bcdot\mathbf{t}\bcdot\frac{\partial\boldsymbol{V}}{\partial x^l}(x,t)
  &&=(-p+\gamma)\theta+\frac{\mu}{2}(\nabla^iV^j+\nabla^jV^i)(\nabla_iV_j+\nabla_jV_i)+q_\Sigma
  \nonumber\\
  &&=:(-p+\gamma)\theta+\frac{\mu}{2}|\boldsymbol{V}\otimes\bnabla+\bnabla\otimes\boldsymbol{V}|^2+q_\Sigma
  \nonumber
\end{eqnarray}
where the last two identities are due to the adoption of the constitutive relation \eqref{Constitutive Relation}.

\section{Vorticity Dynamics}

\subsection{General Theories}

Firstly, the following identity has been derived
\begin{lemma}
\begin{equation}
    \left[\overset{\circ}{\bnabla}\times(\boldsymbol{b}\bcdot\mathbf{F})\right]\bcdot\boldsymbol{N}_\Sigma
   =|\mathbf{F}|\left(\bnabla\times\boldsymbol{b} \right)\bcdot\boldsymbol{n}_\Sigma,\quad
    \forall\,\boldsymbol{b}\in\boldsymbol{T\Sigma},
    \quad |\mathbf{F}|:=\frac{\sqrt{g}}{\sqrt{G}}\det\left[\frac{\partial x^i}{\partial\xi^A}\right](\xi,t)
    \label{Int_Curr_Rel}
\end{equation}
where $\overset{\circ}{\bnabla}\triangleq\boldsymbol{G}^L\overset{\circ}{\nabla}_{\frac{\partial}{\partial\xi^L}}$ and $\bnabla\triangleq\boldsymbol{g}^l\nabla_{\frac{\partial}{\partial x^l}}$ are \emph{\textit{Levi-Civita connection operators}}, $\boldsymbol{N}_\Sigma$ and $\boldsymbol{n}_\Sigma$ are surface normal vectors corresponding to the initial and current physical configurations respectively, $|\mathbf{F}|$ denotes the determinant of $\mathbf{F}$,
$\sqrt{G}:=[\boldsymbol{G}_1,\boldsymbol{G}_2,\boldsymbol{N}_\Sigma]$.
\end{lemma}

\noindent\textbf{Proof}:
\begin{eqnarray}
  && \left[\overset{\circ}{\bnabla}\times(\boldsymbol{b}\bcdot\mathbf{F})\right]\bcdot\boldsymbol{N}_\Sigma
    =\left[ \left(\boldsymbol{G}^B\overset{\circ}{\nabla}_{\frac{\partial}{\partial\xi^B}}\right)\times\left(
     b_i\frac{\partial x^i}{\partial \xi^A}(\xi,t)\boldsymbol{G}^A \right)  \right]\bcdot\boldsymbol{N}_\Sigma
    =\overset{\circ}{\nabla}_B\left(b_i\frac{\partial x^i}{\partial\xi^A}(\xi,t)\epsilon^{BA3} \right)
    \nonumber\\
  &&=\epsilon^{BA3}\overset{\circ}{\nabla}_B\left(b_i\frac{\partial x^i}{\partial\xi^A}(\xi,t)\right)
    =\epsilon^{BA3}\left[ \frac{\partial}{\partial\xi^B}\left(b_i\frac{\partial x^i}{\partial\xi^A} \right)(\xi,t)
    -\Gamma^L_{BA}\left(b_i\frac{\partial x^i}{\partial\xi^L}(\xi,t) \right) \right]
    \nonumber\\
  &&=\epsilon^{BA3} \frac{\partial}{\partial\xi^B}\left(b_i\frac{\partial x^i}{\partial\xi^A} \right)(\xi,t)
    =\frac{\partial b_i}{\partial x^s}(x,t)\left[ \epsilon^{BA3}\frac{\partial x_s}{\partial\xi^B}(\xi,t)
    \frac{\partial x_i}{\partial\xi^A}(\xi,t) \right]
    \nonumber\\
  &&=\frac{1}{\sqrt{G}}\det\left[\frac{\partial x^i}{\partial\xi^A}\right](\xi,t)
     e^{si3}\frac{\partial b_i}{\partial x^s}(x,t)
    =\frac{\sqrt{g}}{\sqrt{G}}\det\left[\frac{\partial x^i}{\partial\xi^A}\right](\xi,t)(\epsilon^{si3}\nabla_sb_i)
    \nonumber\\
  &&=|\mathbf{F}|(\bnabla\times\boldsymbol{b})\bcdot\boldsymbol{n}_\Sigma
  \nonumber
\end{eqnarray}

As an application, the governing equation of vorticity can be derived
\begin{corollary}[Vorticity Equation]
\begin{equation}
    \dot{\omega}^3=-\theta\omega^3+(\bnabla\times\boldsymbol{a})\bcdot\boldsymbol{n}_\Sigma
    \label{Vor_Equ.}
\end{equation}
\end{corollary}

\noindent\textbf{Proof}:
Let $\boldsymbol{b}$ in the relation \eqref{Int_Curr_Rel} be the velocity $\boldsymbol{V}$, it reads
\begin{equation*}
    \omega^3=(\bnabla\times\boldsymbol{V})\bcdot\boldsymbol{n}_\Sigma=\frac{1}{|\mathbf{F}|}
    \left[ \overset{\circ}{\bnabla}\times(\boldsymbol{V}\bcdot\mathbf{F}) \right]\bcdot\boldsymbol{N}_\Sigma
\end{equation*}
Furthermore, one can do the following deduction
\begin{eqnarray}
    \dot{\omega}^3
 &&=-\frac{\theta}{|\mathbf{F}|}
    \left[ \overset{\circ}{\bnabla}\times(\boldsymbol{V}\bcdot\mathbf{F}) \right]\bcdot\boldsymbol{N}_\Sigma
   +\frac{1}{|\mathbf{F}|}
    \left[ \overset{\circ}{\bnabla}\times(\boldsymbol{a}\bcdot\mathbf{F})
          +\overset{\circ}{\bnabla}\times\left(\, \boldsymbol{V}
          \bcdot ( \boldsymbol{V}\otimes\overset{\Sigma}{\bnabla} )\bcdot\mathbf{F} \,\right)
    \right]\bcdot\boldsymbol{N}_\Sigma
    \nonumber\\
 &&=-\theta(\bnabla\times\boldsymbol{V})\bcdot\boldsymbol{n}_\Sigma+(\bnabla\times\boldsymbol{a})\bcdot\boldsymbol{n}_\Sigma
   +\left( \bnabla\times\bnabla\left(\frac{|\boldsymbol{V}|^2}{2}\right) \right)\bcdot\boldsymbol{n}_\Sigma
   =-\theta\omega^3+(\bnabla\times\boldsymbol{a})\bcdot\boldsymbol{n}_\Sigma
    \nonumber
\end{eqnarray}
where the identities $\dot{\mathbf{F}}=( \boldsymbol{V}\otimes\overset{\Sigma}{\bnabla} )\bcdot\mathbf{F}$ and $\mathrm{d}|\mathbf{F}|/\mathrm{d}t=\theta|\mathbf{F}|$ \cite[see][]{XXL-2013} are utilized.

As an appendant of the above proof, one has the conclusion
\begin{corollary}[Lagrange Theorem on Vorticity]
In the case of the acceleration field is irrotational, a patch of continuous medium that is initially irrotational will keep irrotational at any time, conversely a patch of continuous medium with initially nonzero vorticity will possess vorticity at any time although the value can be changed.
\end{corollary}

Secondly, the following identity can be readily set up
\begin{lemma}
\begin{equation}
    \bnabla\times(\bnabla\times\boldsymbol{b})=\bnabla(\bnabla\bcdot\boldsymbol{b})+K_G\boldsymbol{b}
    -\boldsymbol{\Delta}\boldsymbol{b},\quad
    \forall\,\boldsymbol{b}\in\boldsymbol{T\Sigma},
    \quad\boldsymbol{\Delta}\boldsymbol{b}\triangleq\bnabla\bcdot(\bnabla\times\boldsymbol{b})
    \nonumber
\end{equation}
\end{lemma}
\noindent\textbf{Proof}:
\begin{alignat*}{3}
    \bnabla\times(\bnabla\times\boldsymbol{b})
   &=\bnabla\times\left[ \left(\boldsymbol{g}^p\nabla_{\frac{\partial}{\partial x^p}}\right) \times
     \left( b_i\boldsymbol{g}^i\right) \right]
    =\left( \boldsymbol{g}^q\nabla_{\frac{\partial}{\partial x^q}} \right) \times \left[(\nabla_p b_i)\epsilon^{pi3}\boldsymbol{n} \right]\\
   &=\epsilon^{3kq}\epsilon_{3pi}\nabla_q(\nabla^p b^i)\boldsymbol{g}_k
    =(\delta^k_p\delta^q_i-\delta^q_p\delta^k_i)\nabla_q(\nabla^p b^i)\boldsymbol{g}_k
    =\nabla_i(\nabla^k b^i)\boldsymbol{g}_k-\nabla_p(\nabla^p b^i)\boldsymbol{g}_i\\
   &=\nabla_i(\nabla^k b^i)\boldsymbol{g}_k-\boldsymbol{\Delta}\boldsymbol{b}
\end{alignat*}
Furthermore, one has
\begin{alignat*}{3}
    \nabla_i(\nabla^k b^i)\boldsymbol{g}_k
  &=[\nabla^k(\nabla_i b^i)+R^{i\cdot\cdot k}_{\cdot si\cdot} b^s]\boldsymbol{g}_k
   =\bnabla(\bnabla\cdot\boldsymbol{b}) + K_G(\delta^i_i\delta^k_s-g_{si}g^{ik})b^s\boldsymbol{g}_k \\
  &=\bnabla(\bnabla\cdot\boldsymbol{b}) + K_G\boldsymbol{b}
\end{alignat*}
where the change of the order of covariant or contra-variant derivatives should be related to Riemannian-Christoffel tensor. It is the end of the proof. It should be pointed out that this kind of identities is still keeping valid for any tensor field.

As an application, the governing equation of momentum conservation on the tangent plane \eqref{NSE-com} can be rewritten as
\begin{equation}
    \rho\,a_l\boldsymbol{g}^l=\bnabla\Pi-\mu\bnabla\times\boldsymbol{\omega}+2\mu K_G\boldsymbol{V}+f_{sur,l}\boldsymbol{g}^l,
    \quad \Pi:=-p+2\mu\theta
    \label{NSE}
\end{equation}
where $\bnabla\times\boldsymbol{\omega}=\epsilon^{k3l}\frac{\partial\omega^3}{\partial x^k}(x,t)\boldsymbol{g}_l$.
Subsequently, the following coupling relations can be attained just by doing the dot and cross products by $\boldsymbol{e}$ on both sides of \eqref{NSE} respectively.
\begin{corollary}[Coupling Relations between Directional Derivatives of $\Pi$ and $\boldsymbol{\omega}$ ]
\begin{equation*}
    \left\{
    \begin{array}{l}
        \displaystyle
        \frac{\partial\Pi}{\partial\boldsymbol{e}}=\rho\boldsymbol{a}\bcdot\boldsymbol{e}       +\mu(\bnabla\times\boldsymbol{\omega})\bcdot\boldsymbol{e}-2\mu K_G\boldsymbol{V}\bcdot\boldsymbol{e}
        -\boldsymbol{f}_\Sigma\bcdot\boldsymbol{e} \\[12pt]
        \displaystyle
        \mu\frac{\partial\omega^3}{\partial\boldsymbol{e}}=-[\rho\boldsymbol{a},\boldsymbol{e},\boldsymbol{n}_\Sigma]
        +[\bnabla\Pi,\boldsymbol{e},\boldsymbol{n}_\Sigma]+[2\mu K_G\boldsymbol{V},\boldsymbol{e},\boldsymbol{n}_\Sigma]
        +[\boldsymbol{f}_\Sigma,\boldsymbol{e},\boldsymbol{n}_\Sigma]
    \end{array}
    \right.
\end{equation*}
for all $\boldsymbol{e}\in\boldsymbol{T\Sigma}$ s.t $|\boldsymbol{e}|=1$. The coupling relations are valid at any point in the flow field.
\end{corollary}

Thirdly, the intrinsic decomposition is still valid for any surface tensor, i.e. there exists
\begin{lemma}[Intrinsic Decomposition]
\begin{equation*}
    \mathbf{\Phi}=\left\{
    \begin{array}{l}
    \displaystyle
    \boldsymbol{e}\otimes(\boldsymbol{e}\bcdot\,\mathbf{\Phi})-[\,\boldsymbol{e},\,[\boldsymbol{e},\mathbf{\Phi}] \,]\\[16pt]
    \displaystyle
    (\mathbf{\Phi}\bcdot\boldsymbol{e})\otimes\boldsymbol{e}-[\,[\mathbf{\Phi},\boldsymbol{e}],\,\boldsymbol{e} \,]
    \end{array}
    \right.,
    \quad
    \forall\mathbf{\Phi}\in\mathscr{T}^r(\boldsymbol{T\Sigma}),
    \quad\boldsymbol{e}\in\boldsymbol{T\Sigma}\,\,\mbox{s.t.}\,\,|\boldsymbol{e}|=1
\end{equation*}
where $[\boldsymbol{e},\mathbf{\Phi}]$ denotes the cross product $\boldsymbol{e}\times\mathbf{\Phi}$ and so on.
\end{lemma}
\noindent\textbf{Proof}:
Let $\mathbf{\Phi}=\Phi^i_{\cdot j}\boldsymbol{g}_i\otimes\boldsymbol{g}^j\in\mathscr{T}^2(\boldsymbol{T\Sigma})$ without lost of generality, and to calculate
\begin{eqnarray}
    [\, [\mathbf{\Phi},\boldsymbol{e}],\,\boldsymbol{e} \,]
 &&=[\, [\Phi^i_{\cdot j}\boldsymbol{g}_i\otimes\boldsymbol{g}^j, e_k\boldsymbol{g}^k ],\,\boldsymbol{e} \,]
   =[\, \Phi^i_{\cdot j}e_k\epsilon^{jk3}\boldsymbol{g}_i\otimes\boldsymbol{n}_\Sigma,\,\boldsymbol{e} \,]
   =\Phi^i_{\cdot j}(e_k e^q)(\epsilon^{jk3}\epsilon_{ql3})\boldsymbol{g}_i\otimes\boldsymbol{g}^l
   \nonumber \\
 &&=\Phi^i_{\cdot j}(e_k e^q)(\delta^j_q\delta^k_l-\delta^k_q\delta^j_l)\boldsymbol{g}_i\otimes\boldsymbol{g}^l
   =(\Phi^i_{\cdot j} e^j\boldsymbol{g}_i)\otimes(e_k\boldsymbol{g}^k)-\mathbf{\Phi}
   \nonumber
\end{eqnarray}
Evidently, one arrives at the conclusion. The other identity can be proved similarly.

Applied the intrinsic decomposition to the velocity gradient tensor, the strain tensor has the following representation, termed still as Caswell formula.
\begin{corollary}[Caswell Formula]
On any fixed solid boundary where the general viscous boundary condition is satisfied by the fluid, namely $\boldsymbol{V}=0$ on the boundary, the strain tensor can be represented as following
\begin{equation*}
    \mathbf{D}=\theta\boldsymbol{n}\otimes\boldsymbol{n}
    +\frac{1}{2}(\boldsymbol{\omega}\times\boldsymbol{n})\otimes\boldsymbol{n}
    +\frac{1}{2}\boldsymbol{n}\otimes(\boldsymbol{\omega}\times\boldsymbol{n})
\end{equation*}
\end{corollary}

\noindent\textbf{Proof}:
Basically, a kind of local curvilinear coordinates denoted by $\{x^\shortparallel,x^\bot\}$ corresponding to the fixed solid boundary can be set up such that the local co-variant basis $\{ \boldsymbol{g}_\shortparallel,\boldsymbol{g}_\bot \}$ is orthogonal on the boundary with $\boldsymbol{g}_\shortparallel$ tangents to the boundary and $\boldsymbol{g}_\bot=\boldsymbol{g}^\bot=\boldsymbol{n}$. The details can be referred to the appendix.

Subsequently, the intrinsic decomposition is utilized
\begin{equation*}
    \boldsymbol{V}\otimes\bnabla=( \boldsymbol{V}\otimes\bnabla,\boldsymbol{n})\otimes\boldsymbol{n}
    -[\,[\,\boldsymbol{V}\otimes\bnabla,\boldsymbol{n}\,],\,\boldsymbol{n}\,]
\end{equation*}

For the first term on the right hand, one has
\begin{alignat*}{3}
    (\boldsymbol{V}\otimes\bnabla,\boldsymbol{n})
  &=(\boldsymbol{V}\otimes\bnabla-\bnabla\otimes\boldsymbol{V},\boldsymbol{n})+(\bnabla\otimes\boldsymbol{V},\boldsymbol{n})
   =\boldsymbol{\omega}\times\boldsymbol{n}+(\bnabla\otimes\boldsymbol{V},\boldsymbol{n})
\end{alignat*}
where
\begin{alignat*}{3}
    (\bnabla\otimes\boldsymbol{V},\boldsymbol{n})
  &=\left( \boldsymbol{g}^\shortparallel\otimes\bnabla_\frac{\partial}{\partial x^\shortparallel}\boldsymbol{V}
          +\boldsymbol{g}^\bot\otimes\bnabla_\frac{\partial}{\partial x^\bot}\boldsymbol{V}, \boldsymbol{n} \right)
   =\theta\boldsymbol{n}
\end{alignat*}
thanks to the relations
\begin{equation*}
    \bnabla_\frac{\partial}{\partial x^\shortparallel}\boldsymbol{V}
   =\bnabla_\shortparallel v^\shortparallel\boldsymbol{g}_\shortparallel
   +\bnabla_\shortparallel v^\bot\boldsymbol{g}_\bot=0\in\boldsymbol{T\Sigma}
\end{equation*}
due to the viscous boundary condition and
\begin{equation*}
    \left( \bnabla_\frac{\partial}{\partial x^\bot}\boldsymbol{V},\boldsymbol{n} \right)
   =\left( \bnabla_\bot v^\shortparallel\boldsymbol{g}_\shortparallel
   +\bnabla_\bot v^\bot\boldsymbol{g}_\bot,\boldsymbol{g}^\bot \right)
   =\bnabla_\bot v^\bot=\theta-\bnabla_\shortparallel v^\shortparallel=\theta
\end{equation*}

For the second term on the right hand, it is naught thanks to
\begin{equation*}
    [\boldsymbol{V}\otimes\bnabla,\boldsymbol{n}]
   =[\,
        \left(\bnabla_\shortparallel\boldsymbol{V}\right)\otimes\boldsymbol{g}^\shortparallel
       +\left(\bnabla_\bot\boldsymbol{V}\right)\otimes\boldsymbol{g}^\bot,\boldsymbol{n}
    \,]
   =[\, \left(\bnabla_\bot\boldsymbol{V}\right)\otimes\boldsymbol{g}^\bot,\boldsymbol{g}^\bot \,]
   =\left(\bnabla_\bot\boldsymbol{V}\right)\otimes[\,\boldsymbol{g}^\bot,\boldsymbol{g}^\bot \,]
   =0
\end{equation*}

As a summary, one arrives at the relation
\begin{equation*}
    \boldsymbol{V}\otimes\bnabla=(\boldsymbol{\omega}\times\boldsymbol{n})\otimes\boldsymbol{n}
    +\theta\boldsymbol{n}\otimes\boldsymbol{n}
\end{equation*}
Accompanying with its conjugate relation, the proof is completed.

Readily, one has the following assertion
\begin{corollary}
    For any two dimensional incompressible viscous flow on any fixed smooth surface, on any fixed solid boundary, the directions corresponding to the maximum or minimum rate of change of element material arc length with the same absolute value $|\omega^3|/2$ is $\pi/4$ or $3\pi/4$ with respect to the tangent direction of the boundary.
\end{corollary}

\subsection{Particular Theories For Incompressible Flows}

As generally, the incompressibility for two dimensional flows on fixed smooth surfaces is still defined as $\dot{\theta}=0$. \cite{Chomaz-1990} revealed some velocity domains
of soap films in which they can be considered as two dimensional incompressible flows. Based on the continuity equation \eqref{MassCon_Diff}, one has
\begin{equation*}
    \theta=\nabla_s V^s=\frac{\partial V^s}{\partial x^s}(x,t)+\Gamma^s_{sl}V^l
    =\frac{1}{\sqrt{g}}\frac{\partial}{\partial x^s} (\sqrt{g}V^s)(x,t)=0
\end{equation*}
then the stream function can be introduced through $V^s=\epsilon^{st3}\frac{\partial\psi}{\partial x^t}(x,t)$. Subsequently, the stream function \& vorticity algorithm can be derived.
\begin{lemma}[Stream Function \& Vorticity Algorithm for Incompressible Flows]
\begin{equation*}
\left\{
\begin{array}{l}
\displaystyle
   \triangle\psi\triangleq g^{ij}\left[\frac{\partial^2\psi}{\partial x^i\partial x^j}(x,t)-\Gamma^k_{ij}\frac{\partial\psi}{\partial x^k}(x,t) \right]=-\omega^3 \\[12pt]
\displaystyle
    \dot{\omega}^3
   =\frac{\partial \omega^3}{\partial t}(x,t)+V^s\frac{\partial \omega^3}{\partial x^s}(x,t)
   =\frac{\mu}{\rho}\left[\, \nabla^s\nabla_s\omega^3+2\epsilon^{kl3}\nabla_k(K_G V_l) \,\right]
   +\frac{1}{\rho}\epsilon^{kl3}\nabla_k f_{sur,l}
\end{array}
\right.
\end{equation*}
\end{lemma}
\noindent\textbf{Proof}:
The Stream function Possion equation is just from the definition $\omega^3=\epsilon_{3ts}\nabla^tV^s$. On vorticity equation, accompanying \eqref{Vor_Equ.} with \eqref{NSE-com} and taking account of $\theta=0$, one has
\begin{eqnarray}
   \dot{\omega}^3
  &&=(\bnabla\times\boldsymbol{a})\bcdot\boldsymbol{n}_\Sigma
    =\epsilon^{kl3}\nabla_k a_l
  \nonumber \\
  &&=-\frac{1}{\rho}\epsilon^{kl3}\frac{\partial^2 p}{\partial x^kx^l}(x,t)
    +\frac{\mu}{\rho}\left[\, \epsilon^{kl3}\nabla_k(\nabla^s\nabla_s V_l) +\epsilon^{kl3}\nabla_k(K_GV_l) \,\right]
    +\frac{1}{\rho}\epsilon^{kl3}\nabla_k f_{sur,l}
  \nonumber
\end{eqnarray}
Thanks to the relation \cite[see][]{XXL-2013}
\begin{equation*}
    \epsilon^{kl3}\nabla_k(\nabla^s\nabla_s V_l)=\nabla^s\nabla_s(\epsilon^{kl3}\nabla_k V_l)
    +\epsilon^{kl3}\nabla_k(K_G V_l)
    =\nabla^s\nabla_s\omega^3+\epsilon^{kl3}\nabla_k(K_G V_l)
\end{equation*}
the proof is completed.

Finally, the pressure Possion equation can be derived
\begin{lemma}[Pressure Possion Equation for Incompressible Flows]
\begin{equation*}
    -\boldsymbol{\Delta} p=\rho [ (\boldsymbol{V}\otimes\bnabla)\boldsymbol{:}(\bnabla\otimes\boldsymbol{V})+K_G|\boldsymbol{V}|^2 ]
    -2\mu\boldsymbol{V}\bcdot(\nabla K_G)-\bnabla\bcdot\boldsymbol{f}_\Sigma,
    \quad |\boldsymbol{V}|^2:=V^sV_s
\end{equation*}
\end{lemma}

\noindent\textbf{Proof}:
Taking denoted as $\nabla\cdot$ the divergence operator on both sides of \eqref{NSE-com}, one has
\begin{equation*}
    \rho\nabla^l\left( \frac{\partial V_l}{\partial t}(x,t)+V^s\nabla_s V_l \right)
   =-\nabla^l(\nabla_l p)
   +\mu\left[ \nabla^l\nabla^s\nabla_s V_l +\nabla^l(K_G V_l) \right]+\nabla^l f_{sur,l}
\end{equation*}
where
\begin{eqnarray}
    \nabla^l(V^s\nabla_s V_l)
 &&=(\nabla^lV^s)(\nabla_s V_l)+V^s(\nabla^l\nabla_sV_l)
   =(\nabla^lV^s)(\nabla_s V_l)+V^s\left[ \nabla_s(\nabla^lV_l)+R^{\cdot tl \cdot}_{l\cdot\cdot s}V_t \right]
   \nonumber \\
 &&=(\nabla^lV^s)(\nabla_s V_l)+K_G(\delta^l_l\delta^t_s-g^{tl}g_{ls})V^sV_l
   =(\boldsymbol{V}\otimes\nabla):(\boldsymbol{V}\otimes\nabla)+K_G|\boldsymbol{V}|^2
   \nonumber
\end{eqnarray}
\begin{eqnarray}
    \nabla^l(\nabla^s\nabla_s V_l)
 &&=\nabla^l\nabla^s(\nabla_s V_l)
   =\nabla^s\nabla^l(\nabla_sV_l)+R^{\cdot tls}_{s}\nabla_tV_l+R^{\cdot tls}_{l}\nabla_sV_t
   \nonumber \\
 &&=\nabla^s\nabla^l(\nabla_sV_l)+K_G(\delta^l_sg^{ts}-g^{tl}\delta^s_s)\nabla_tV_l
   +K_G(\delta^l_lg^{ts}-g^{tl}\delta^s_l)\nabla_sV_t
   =\nabla^s(\nabla^l\nabla_sV_l)
   \nonumber \\
 &&=\nabla^s [ \nabla_s(\nabla^lV_l) +R^{\cdot tl\cdot}_{l\cdot\cdot s}V_t ]
   =\nabla^s [ K_G(\delta^l_l\delta^t_s-g^{tl}g_{ls})V_t ]=\nabla^s(K_GV_s)=V_s\nabla^sK_G
   \nonumber
\end{eqnarray}
Then the identity is proved.

\section{Conclusions}

Basically, some primary relations in general theory of vorticity dynamics have been extended to surface tensor fields
based on Levi-Civita connection operator. Subsequently, the governing equation of vorticity, Lagrange
theorem on vorticity, Caswell formula on strain tensor are attained. Furthermore, the stream function \& vorticity algorithm with pressure Possion equation has been set up for incompressible flows. It has been revealed that Gaussian and mean curvatures are taking part directly in some governing equations but all of the effects due to the geometry of the surface will disappear as flows on general fixed surfaces degenerates to ones on planes. All of the theoretical results presented in this paper are exactly deduced without any approximation and constitute a theoretical framework of vorticity dynamics for two dimensional flows on fixed smooth surfaces.

\section*{Acknowledgements}

This work is supported by National Nature Science Foundation of China (Grant No.$11172069$) and some key projects of education reforms issued by the Shanghai Municipal Education Commission ($2011$).

\section*{Appendix 1. Construction of local curvilinear coordinates $\{x^\shortparallel,x^\bot\}$ }

\begin{figure}
\begin{center}
    \includegraphics[height=60mm]{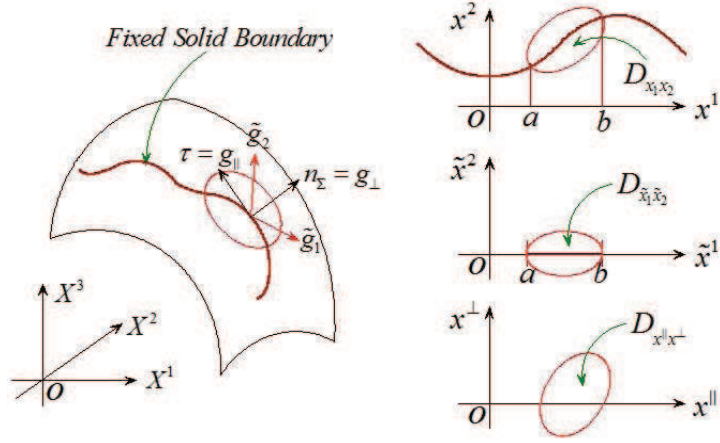}
\end{center}
\caption { Sketch of local parametrization of a patch of surface, $D_{x^1x^2}$, $D_{\tilde{x}^1\tilde{x}^2}$ and $D_{x^\shortparallel x^\bot}$ denote parametric domains with respect to coordinates $\{x^1, x^2\}$, $\{\tilde{x}^1, \tilde{x}^2\}$ and $\{ x^\shortparallel, x^\bot \}$ respectively. }
\label{Draw2}
\end{figure}

As sketched in Figure \ref{Draw2}, one can suppose without lost of generality that the preimage of a segment of boundary on the surface in the parametric space $\{x^1,x^2\}$ can be represented locally as
\begin{equation*}
    (a,b)\ni x^1\,\mapsto\,
    \left[ \begin{array}{c} x^1\\ x^2 \end{array} \right](x^1)
   =\left[ \begin{array}{c} x^1\\ \psi(x^1) \end{array} \right]
\end{equation*}
Then the following coordinates can be constructed locally
\begin{equation*}
    \left[ \begin{array}{c} x^1\\ x^2 \end{array} \right]
    \,\mapsto\,
    \left[ \begin{array}{c} \tilde{x}^1\\ \tilde{x}^2 \end{array} \right](x^1,x^2)
   =\left[ \begin{array}{c} x^1\\ x^2-\psi(x^1) \end{array} \right]
\end{equation*}
The segment of the boundary is corresponding to $\tilde{x}^1\in(a,b)$ and $\tilde{x}^2=0$. The local co-variant basis with respect to $\{\tilde{x}^i\}^2_{i=1}$ is denoted by $\{ \tilde{\boldsymbol{g}}_i\}^2_{i=1}$. Thirdly, one introduces another coordinates $\{x^\shortparallel,x^\bot \}$. Its local co-variant basis $\{ \boldsymbol{g}_\shortparallel,\boldsymbol{g}_\bot\}$
has the following relations with $\{ \tilde{\boldsymbol{g}}_i\}^2_{i=1}$
\begin{equation*}
    [\boldsymbol{g}_\shortparallel,\boldsymbol{g}_\bot]=
    [\tilde{\boldsymbol{g}}_1,\tilde{\boldsymbol{g}_2}]
    \left[
    \begin{array}{cc}
        \displaystyle\frac{\partial\tilde{x}^1}{\partial x^\shortparallel} &  \displaystyle\frac{\partial\tilde{x}^1}{\partial x^\bot}\\[12pt]
        \displaystyle\frac{\partial\tilde{x}^2}{\partial x^\shortparallel} &
        \displaystyle\frac{\partial\tilde{x}^2}{\partial x^\bot}
    \end{array}
    \right](x^\shortparallel,x^\bot )
\end{equation*}
Accompanying the construction of $\{ \boldsymbol{g}_\shortparallel,\boldsymbol{g}_\bot\}$ on the boundary with the general relations
\begin{equation*}
    \left[
    \begin{array}{cc}
        \displaystyle\frac{\partial x^\shortparallel}{\partial\tilde{x}^1}&
        \displaystyle\frac{\partial x^\shortparallel}{\partial\tilde{x}^2}\\[12pt]
        \displaystyle\frac{\partial x^\bot}{\partial\tilde{x}^1}&
        \displaystyle\frac{\partial x^\bot}{\partial\tilde{x}^2}\\
      \end{array}
    \right](\tilde{x}^1,\tilde{x}^2 )
    =
    \left[
    \begin{array}{cc}
        \displaystyle\frac{\partial\tilde{x}^1}{\partial x^\shortparallel} &
        \displaystyle\frac{\partial\tilde{x}^1}{\partial x^\bot}\\[12pt]
        \displaystyle\frac{\partial\tilde{x}^2}{\partial x^\shortparallel} &
        \displaystyle\frac{\partial\tilde{x}^2}{\partial x^\bot}
    \end{array}
    \right]^{-1}(x^\shortparallel,x^\bot )
\end{equation*}
the following Jacobian matrix on the boundary can be determined
\begin{equation*}
    \left[
    \begin{array}{cc}
        \displaystyle\frac{\partial x^\shortparallel}{\partial\tilde{x}^1}& \displaystyle\frac{\partial x^\shortparallel}{\partial\tilde{x}^2}\\[12pt]
        \displaystyle\frac{\partial x^\bot}{\partial\tilde{x}^1}& \displaystyle\frac{\partial x^\bot}{\partial\tilde{x}^2}\\
      \end{array}
    \right](\tilde{x}^1,0),\quad\tilde{x}^1\in(a,b)
\end{equation*}
then the coordinates $\{x^\shortparallel,x^\bot \}$ can be constructed as follows
\begin{equation*}
\left\{
\begin{array}{l}
\displaystyle
     x^\shortparallel(\tilde{x}^1,\tilde{x}^2) = \int \frac{\partial x^\shortparallel}{\partial\tilde{x}^1}(\tilde{x}^1,0)\,d\tilde{x}^1
     + \frac{\partial x^\shortparallel}{\partial\tilde{x}^2}(\tilde{x}^1,0)\cdot\tilde{x}^2 \\[12pt]
\displaystyle
     x^\bot(\tilde{x}^1,\tilde{x}^2) = \int \frac{\partial x^\bot}{\partial\tilde{x}^1}(\tilde{x}^1,0)\,d\tilde{x}^1
     + \frac{\partial x^\bot}{\partial\tilde{x}^2}(\tilde{x}^1,0)\cdot\tilde{x}^2\\
\end{array}
\right.
\end{equation*}

\end{document}